\documentclass[
reprint,
superscriptaddress,
amsmath,amssymb,
aps,
prb
]{revtex4-2}

\usepackage{graphicx}% Include figure files
\usepackage{bm}% bold math
\usepackage{sidecap}
\usepackage{xcolor}
\usepackage[colorlinks=true,citecolor=blue%,
]{hyperref}
\usepackage{natbib}

\makeatletter

\newcommand{\Rmnum}[1]{\expandafter\@slowromancap\romannumeral #1@}
\makeatother

\begin{document}

	\title{Port Reconfigurable Phase-Change Optical Resonator}% Force line breaks 
	
	\author{Haiyu Meng}
	\affiliation{Department of Physics, National University of Singapore, Singapore 117542}
	\affiliation{Science, Mathematics and Technology (SMT), Singapore University of Technology and Design, Singapore 497372}
	\affiliation{Center for High-Resolution Electron Microscopy, College of Materials Science and Engineering, Hunan University, Changsha 410082, China}
	
	\author{Lingling Wang}
	\affiliation{School of Physics and Electronics, Hunan University, Changsha 410082, China}
	
	\author{Ziran Liu}
	\affiliation{Department of Physics, Key Laboratory for Low-Dimensional Structures and Quantum Manipulation (Ministry of Education), Hunan Normal University, Changsha 410081, China}
	
	\author{Jianghua Chen}
	\email{jhchen123@hnu.edu.cn}
	\affiliation{Center for High-Resolution Electron Microscopy, College of Materials Science and Engineering, Hunan University, Changsha 410082, China}
	
	\author{Ching Hua Lee}
	\email{phylch@nus.edu.sg}
	\affiliation{Department of Physics, National University of Singapore, Singapore 117542}
	
	\author{Yee Sin Ang}
	\email{yeesin\_ang@sutd.edu.sg}
	\affiliation{Science, Mathematics and Technology (SMT), Singapore University of Technology and Design, Singapore 497372}

	\begin{abstract}
		Active control and manipulation of electromagnetic waves are highly desirable for advanced photonic device technology, such as optical cloaking, active camouflage and information processing. Designing optical resonators with high ease-of-control and reconfigurability remains a open challenge thus far. Here we propose a novel mechanism to continuously reconfigure an optical resonator between one-port and two-port configurations via \emph{phase-change material} for efficient optical modulation. By incorporating a phase-change material VO$_2$ substrate into a photonic crystal optical resonator, we computationally show that the system behaves as a one-port device with near-perfect absorption and two-port device with high transmission up to 92\% when VO$_2$ is in the metallic rutile phase and insulating monoclinic phase, respectively. The optical response can be continuously and reversibly modulated between various intermediate states. More importantly, the proposed device is compatible with wide-angle operation and is robust against structural distortion. Our findings reveal a novel device architecture of \emph{port reconfigurable} optical resonator uniquely enabled by switchable optical properties of phase change material.
	\end{abstract}
	\maketitle
	
	%\begin{introduction}
	Efficient active controlling of electromagnetic waves represent one of the centerpieces of modern photonic device technology \cite{zhang2012controlling,baranov2017coherent,li2020tunable}. Multiple strategies have been extensively explored in recent years to achieve active control of electromagnetic wave transmission and absorption, including the optical interference \cite{chen2012interference} in metamaterials, electromagnetically induced transparency \cite{yang2014all} in an opaque medium, the coherent control of multiple incident light \cite{fan2015tunable,yan2018coherent} and the exploiting critical coupling \cite{fan2003temporal} between light and a resonator. Although excellent absorption or transmission efficiency has been demonstrated in various optical device architectures, the practical design of a single optical device that can be efficiently and dynamically switched between transmission and absorption modes using convenient tuning knobs of good ease-of-control, such as electrical voltage or temperature, remains an ongoing challenge thus far.
	
	Optical resonator is commonly used for electromagnetic waves manipulation. The critical coupling effect, at which the incident light is maximally absorbed when the radiative rate of the energy output from the resonator equals to the dissipative loss rate, offers a practical route to achieve perfect absorption \cite{yariv2002critical,horng2020perfect,wang2019manipulating,kim2017electrical,botten1997periodic}. The theoretical maximum absorption efficiency, $A_{max}$, of an optical resonator is closely related to its port number configuration \cite{wang2019manipulating}. An optical resonator under the one-port configuration [see Fig. 1(b)] can be constructed by placing a metal back mirror on the transmitted side to completely shut down the output port of the two-port optical resonator. Such configuration can produce maximum absorption \cite{liu2014approaching,manceau2013optical,piper2014total} when the critical coupling condition is met, and an exceptional absorption of 95\% has been experimentally demonstrated \cite{wang2015enhanced}. The one-port configuration is, however, fundamentally incompatible with the transmission mode operation where a two-port configuration [see Fig. 1(a)] is needed to transmit the electromagnetic wave \cite{wang2019manipulating,xiao2020tailoring}. 
	
	\begin{SCfigure*}
		\caption{\label{Fig:jhhhhhpg} Concept of the reconfigurable phase-change material optical resonator device. Schematic drawings of (a) Two-port optical resonator; (b) One-port optical resonator; and (c) phase-change material (PCM) optical resonator. When the PCM slab is in insulating (metallic) state, the device operates in the two-port (one-port) configuration. (d) Proof-of-concept reconfigurable optical resonator device composed of a photonic crystal slab and a VO$_2$ back slab. (e) The side and the top views of the simulated proof-of-concept device.}
		\includegraphics[scale = 0.425]{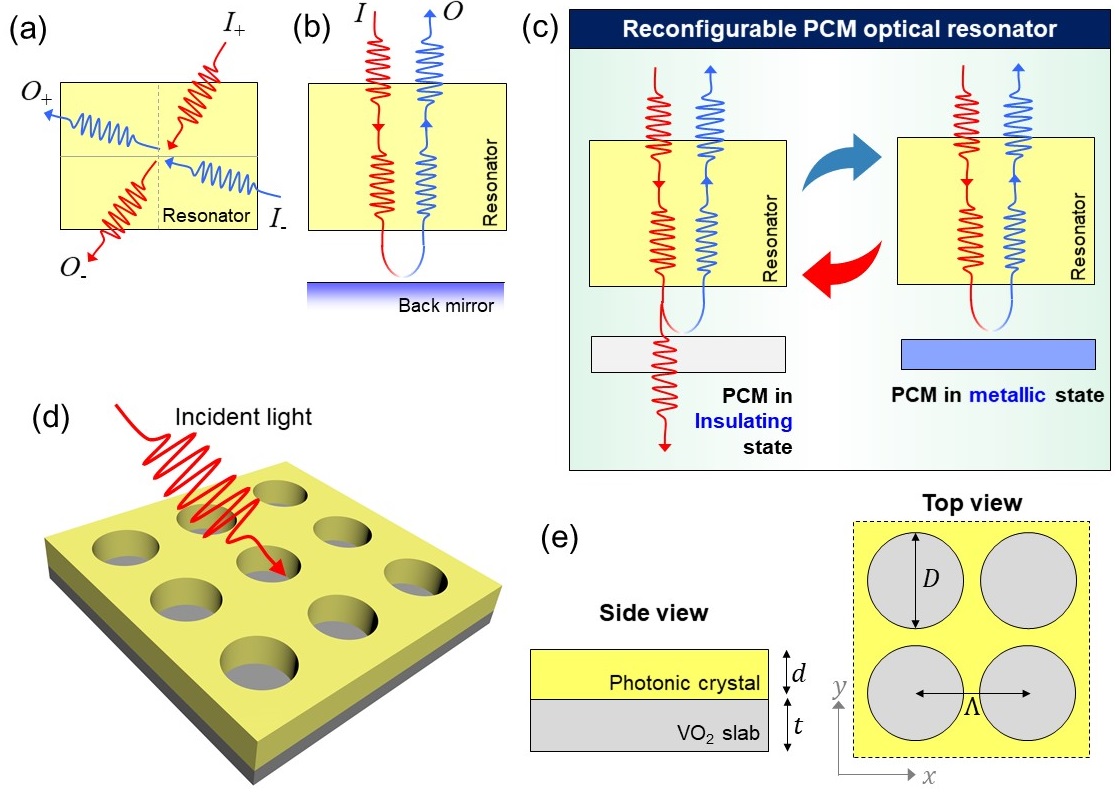}
	\end{SCfigure*}
	
	Such incompatibility immediately raises the following key questions: Can an optical resonator device be dynamically switched between one-port and two-port configurations? If so, how can such port switching mechanism be harnessed to achieve dynamically reconfigurable functions in an optical resonator device? Can the reconfigurable optical resonator be deployed as a robust device that are insensitive to moderate variations of device operation conditions, such as wide-angle of incidence and structural defects? 
	
	In this work, we address these questions by showing that the dynamical closing and opening of the output port offers a strategy to achieve \emph{port reconfigurable} optical resonator that operates between absorption and transmission modes. We propose that such port-closing and port-opening mechanism can be facilitated by a \emph{phase-change material} (PCM) back mirror whose optical properties can be dynamically switched. When the PCM is set into the high-conductivity metallic state, the output port is completely closed, and the optical resonator operates in the one-port mode where strong absorption can be achieved via the critical coupling effect. In contrary, when the PCM is set into low-conductivity insulating state, the PCM back mirror becomes transparent to the electromagnetic waves and the output port is open. The switching between high- and low-conductivity states of the PCM back mirror thus reconfigures the optical resonator between the one-port absorption mode and the two-port transmission mode, respectively. 
	
	In our proof-of-concept device simulation, VO$_2$, a widely studied PCM \cite{driscoll2009memory,seo2010active,wang2015switchable}
	%[cite Science, 2009, 325, 1518., https://pubs.acs.org/doi/10.1021/nl1002153, and https://www.nature.com/articles/srep15020], 
	is used as the PCM back mirror. When combined with a photonic crystal slab optical resonator consisting of silicon, we demonstrate that the transmission intensity is dramatically modulated from 0 to 92\% as the system is switched from one-port to two-port configurations. The optical response of the device is robust against structural imperfection and non-normal incidence, thus revealing the practicality of the proposed device. Unlike typical optical resonators that are limited to a fixed port, our proposed method utilizes dynamically and reversibly reconfigurable port configurations to achieve efficient electromagnetic wave modulation. These findings reveal a radically different approach for achieving ease-of-control optical switching and modulation uniquely enabled by the switchable optical properties of PCM. The proposed port reconfigurable optical resonator shall offer a useful addition to the ever-expanding PCM photonics device family \cite{pitchappa2019chalcogenide, simpson2021phase, gong2021phase}.

	\section{\label{sec:level1}Theory and Computational Model}
	
	\subsection{Theoretical background}
	
	We first describe two well-established approaches, and show how our new approach can overcome their individual shortcomings. For a symmetric two-port system supporting a single resonance, the theoretical maximum of absorption is limited to only
	50\% \cite{thongrattanasiri2012complete,zhang2014coherent}, which can be understood through the symmetric view that an incident wave is decomposed into the superposition of even and odd modes. To overcome this absorption limit, two common methods have been commonly employed: (i) The coherent perfect absorption using two incident beams [Fig. 1(a)] and; (ii) Introducing a perfect back (metal) reflector which effectively closes the output channel of the system and behaves as a one-port configuration [Fig. 1(b)]. While both of these approaches are well-established and have been separately implemented in different devices, our work seeks to combine them in one single photonic setup that can be switched to either as quickly as desired. For method (i), a thin slab can be illuminated from both sides by two beams, denoted here as $I_+$ and $I_-$. Defining the output beams as $O_+$ and $O_-$, thus using scattering matrix, the relationship between input and output is 
	\begin{equation}
		\begin{array}{lcl}
			\begin{bmatrix}
				O_+ \\
				O_-
			\end{bmatrix}
			& = & S\begin{bmatrix}
				I_+ \\
				I_-
			\end{bmatrix}=\begin{bmatrix}
				r_+ & t_-\\
				t_+ & r_-
			\end{bmatrix}\begin{bmatrix}
				I_+ \\
				I_-
			\end{bmatrix}
		\end{array}
	\end{equation}
	In a symmetric lossless system, due to symmetry and reciprocity, the matrix elements can be written as $r_+=r$, $r_-=r$ and $t_+=t$, $t_-=t$. Consider the case where the amplitudes of the two counterpropagating input waves are equal and the phase difference is $\Delta\varphi$, the ratio of the output wave intensities to that of the input wave intensities can be written as
	\begin{equation}
		\eta=\frac{|O_+|^2+|O_-|^2}{|I_+|^2+|I_-|^2}=     \frac{|rI_-e^{i\Delta\varphi}+tI_-|^2+|tI_-e^{i\Delta\varphi}+rI_-|^2}{|I_-e^{i\Delta\varphi}|^2+|I_-|^2}
	\end{equation}
	By simplifying $\eta$, we obtained
	\begin{equation}
		\eta=|t^2+r^2+2rt\cos\Delta\varphi|
	\end{equation}
	The coherent absorptivity is given by $A_{coh}=1-\eta$. Clearly, perfect absorption (i.e. $A_{coh}=1$) can be achieved when $\eta$ = 0. This implies that $|t^2+r^2+2rtcos\Delta\varphi| = 0$, which can be met when \emph{r} = $\pm$ \emph{t} and cos$\Delta\varphi = \mp1$. Thus, a proper phase modulation of the input coherent beams is necessary to achieve the destructive interference of the outgoing waves for perfect absorption \cite{fan2015tunable,espinosa2018coherent}. Furthermore, the absorption and transmission intensities can be continuously tuned by modulating the relative phases of the input beams. 
	For method (ii), the addition of the back reflector overcomes the 50\% limit by transforming the two-port system into an asymmetrical one-port system where the optical resonator can only be accessed by the incident wave from one side. In this scenario, based on a detailed coupled mode theory (CMT) analysis (see Appendix A for a detailed derivation analysis of the CMT for two-port and one-port system), the absorption of such a one-port system can be obtained as $A = 1-\Gamma^2$, 
	\begin{equation}
		\Gamma = \frac{-j(\omega-\omega_0)+\gamma-\delta}{j(\omega-\omega_0)+\gamma+\delta}
	\end{equation}
	where $\omega$ is the frequency of the incident wave, $\omega_0$ is the resonance frequency of the resonator, $\gamma$ is the radiative decay rate constant and $\delta$ is the energy dissipation rate constant. Here the absorption efficiency reaches a possible maximal 100\% when the system is driven at the resonance frequency $\omega=\omega_0$ and the rate constants meet the requirement of $\gamma=\delta$. In this case, benefiting from the existence of back mirror, the transmitted wave is completely suppressed; thus the front side of the resonator becomes the only coupled port between external excitation and the resonator. The normal incidence wave will be reflected back along the same path [see Fig. 1(b)]. Critical coupling occurs in which the reflected wave at the front of the film is exactly canceled out and all the incident light power is limited into the resonator when parameters are chosen as $\gamma=\delta$. 
	
	Although methods (i) and (ii) can be employed \emph{separately} to achieve strong absorption or transmission, they cannot be dynamically switched between these two operation modes as the port configuration of a typical resonator is fixed. In the following, we show that this constraint can be relaxed by synergizing the switchable optical properties of a PCM slab with an optical resonator. The switching of PCM between metallic and insulating states dynamically reconfigures the optical resonator between one-port and two-port configurations, thus allowing both transmission and absorption operation modes to be covered in a single device.
	
	\begin{SCfigure*}
		\caption{\label{Fig:jpg} (a) Calculated absorption spectra at metallic state and transmission spectra at insulating state, respectively. Dual-band absorption spectra are obtained by analytical calculations (red star) based on Eq. (A27) in Appendix A and the data retrieved from FDTD-simulated (blue balls). Geometrical parameters are optimized as $\Lambda$ = 100 $\mu$m, \emph{d} = 15.5 $\mu$m, and air hole radius \emph{r} = 0.31$\Lambda$ = 31 $\mu$m. (b) Absorption spectra at normal incidence under TE (hollow circles) and TM (solid line) polarizations, respectively. Distribution of $|Ez|^2$ in the xy plane for two guided resonance frequencies located at 1.7 THz (c) and 2.33 THz (d).}
		\includegraphics[scale=0.4158]{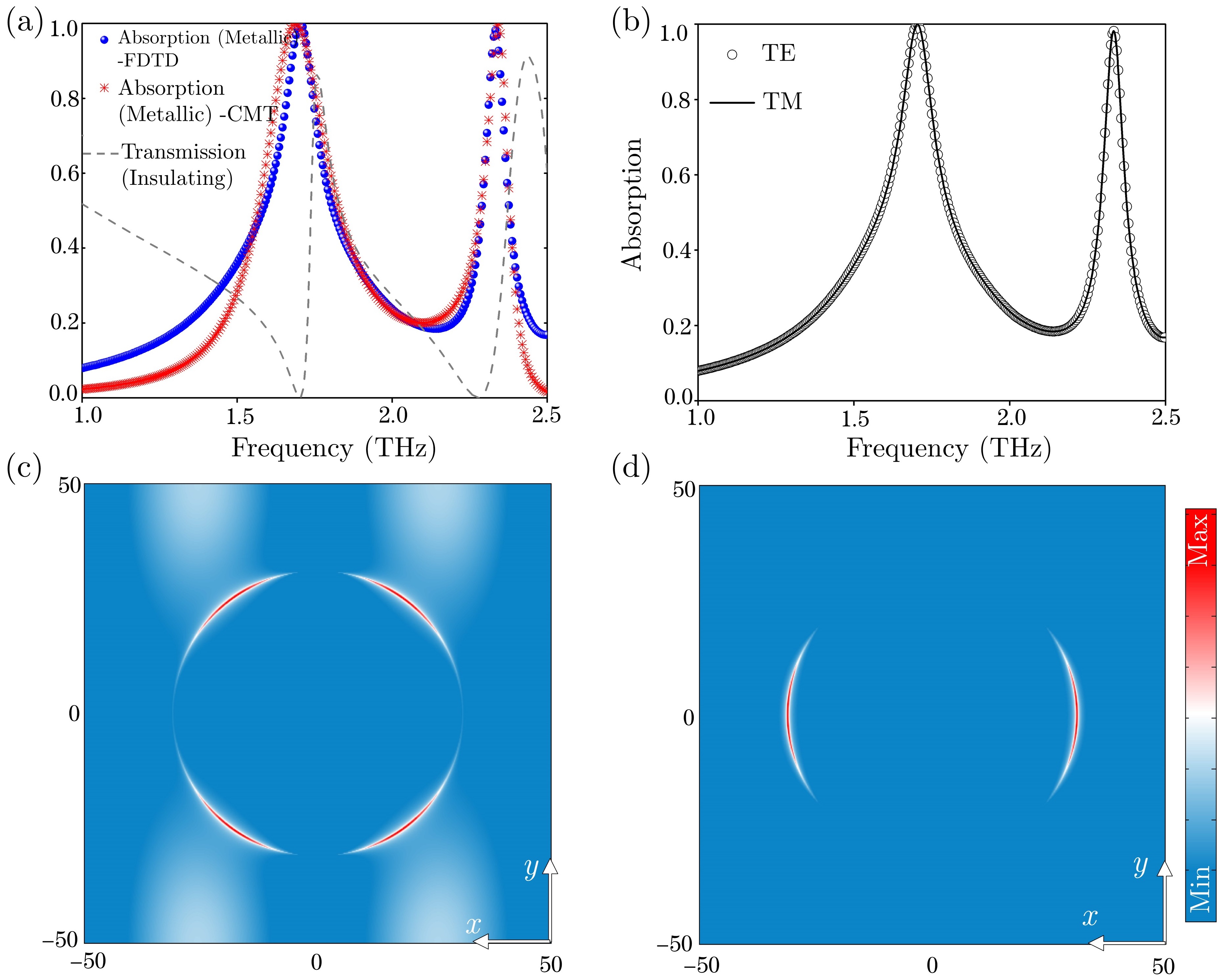}
	\end{SCfigure*}
	
	\subsection{Computational design of port reconfigurable optical resonator}
	
	We consider a photonic crystal slab as the optical resonator with an VO$_2$ substrate serving as the PCM back mirror [see Fig. 1(d)]. Actually, photonic crystal has been employed for characterizing full-colour displays \cite{arsenault2007photonic}, modulator \cite{li2020lithium} and is also ideal platform for the slow light \cite{zhang2019entangled,lin2017line}. Here VO$_2$ is chosen as the candidate PCM due to its dramatic insulator-metal transition (IMT) between the insulating monoclinic state to the metallic rutile state at the transition temperature of $\sim67^\circ$ \cite{stefanovich2000electrical,wu2011electric,nakano2012collective,liu2012terahertz}. The electrical conductivity changes by approximately 5 orders of magnitude during the IMT, resulting in a strong reconfiguration of its optical absorption and dielectric function \cite{wang2017hybrid,kim2019phase}. In addition to direct heating, the phase transition of VO$_2$ can be induced via strain \cite{becker1994femtosecond}, chemical doping \cite{tan2012unraveling}, light \cite{cavalleri2001femtosecond}, electrostatic fields \cite{wu2011electric,nakano2012collective} and electrical voltage \cite{bae2013memristive}, thus, greatly enhancing the ease-of-control VO$_2$-based active optical devices, such as THz waves modulator \cite{zhang2019active}, polaritonic absorber \cite{peng2020tunable}, active terahertz nano-antennas \cite{seo2010active}, and tunable phase modulator \cite{kim2019phase}.
	
	The proposed device is in single-port or two-port operation when the VO$_2$ is in metallic state or insulating state, respectively. A photonic crystal slab ($\varepsilon=12.1$) composed of a sheet of silicon with a periodic array of cylindrical air holes arranged in a square lattice is employed as the optical resonator that can support guided resonances [see Figs. 1(d) and (e)]. The photonic crystal is characterized by lattice period $\Lambda$, thickness $d$, and the air hole radius $r$. The guided mode in photonic crystal slabs interacts with the internal radiation under phase-matched coupling, thus generating the leaky guided resonance \cite{fan2002analysis,zhou2019multiple}.

	In the THz region, the complex permittivity of VO$_2$ can be obtained from the Drude model as
	\begin{equation}
		\epsilon_{VO_2}(\omega)=\epsilon_\infty-\frac{(\omega_p(\sigma_{VO_2)})^2}{\omega^2+i\gamma'\omega}
	\end{equation}
	with $\omega_p(\sigma_{VO_2})=1.4\times10^{15}$ rad/s is the conductivity dependent plasmon frequency, $\gamma'$ is the carrier collision frequency
	\cite{wang2017hybrid,song2019terahertz}. 
	The VO$_2$ changes from an insulator to a fully metallic state when the conductivity switches from 10 to 270000 S/m. Such IMT has been previously demonstrated experimentally using a large variety of tuning knobs \cite{liu2017metamaterials} 
	%\textcolor{red}{Cite https://iopscience.iop.org/article/10.1088/1361-6528/aa9cb1/pdf]}, 
	including thermal \cite{kizuka2015temperature,hilton2007enhanced},
	%\textcolor{red}{Cite https://journals.aps.org/prl/pdf/10.1103/PhysRevLett.99.226401}, 
	light \cite{cavalleri2001femtosecond} and voltage controls
	\cite{joushaghani2014voltage,chen2019gate,zhouvoltage}. The system is illuminated from above by a plane wave with unitary electric field amplitude ($|E_0|=1$) at various incidence angle $\varTheta_0$ and at the few THz frequency regimes. As demonstrated in the following, the optical response of the system can be dynamically modified through the phase transition of the VO$_2$ substrate, thus enabling the dynamical reconfigurable operation of the phase-change optical resonator.

	\section{Results and Discussions}
	
	\subsubsection{Absorption and transmission properties in metallic and insulating VO$_2$ phase}
	
	Figure 2(a) shows the absorption and transmission spectra of the structure under TM-polarized incidence with VO$_2$ in the fully metallic state (blue balls) and fully insulating state (dashed line), respectively. 
	%A TM-polarized wave has its electric field parallel to the incidence plane.
	In this configuration, the electric field is polarized along \emph{x} direction [see device top view in Fig. 1(e)]. 
	%The photonic crystal slab has a periodicity $\Lambda$ = 100 $\mu$m, thickness $d$ = 15.5 $\mu$m, and air hole radius $r$ = 0.31$\Lambda$ = 31 $\mu$m. 
	When VO$_2$ is in the fully metallic phase, the VO$_2$ slab acts as a \emph{fully reflective back mirror}, and the system operates effectively as a \emph{one-port system}. The absorption spectra approach 100\% at the resonance frequency 1.70 and 2.33 THz, indicating that all the incident power is absorbed via the critical coupling effect. 
	In contrast, when VO$_2$ is in the low-conductivity insulating state, the VO$_2$ slab becomes a highly \emph{transmittive} back ‘glass’ that facilitates the transmission of an incident wave, and the device operates in the two-port mode with transmission sharp peaks at 1.75 and 2.43 THz [Fig. 2(a)].

	\begin{SCfigure*}
		\includegraphics[scale=0.2058]{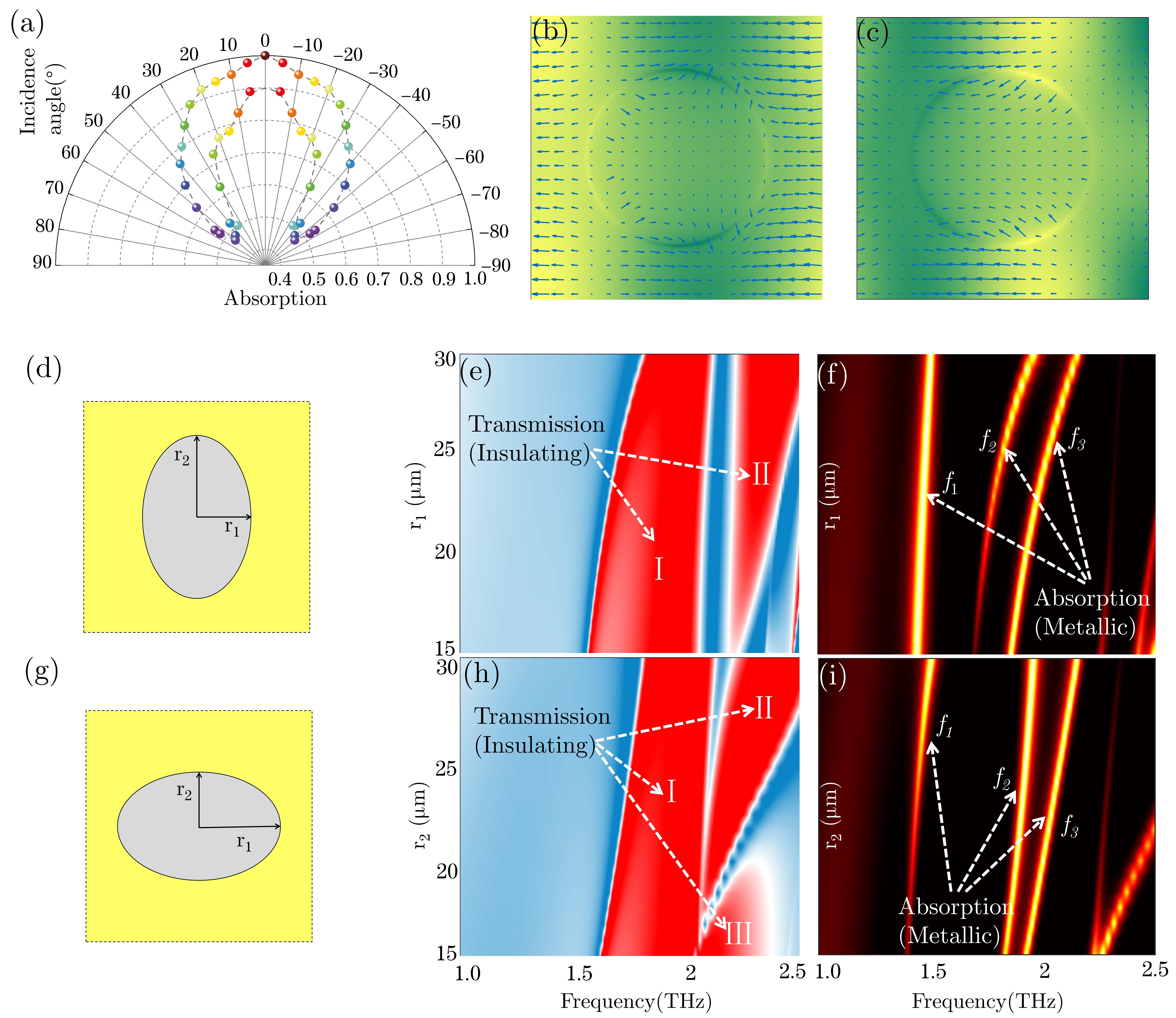}
		\caption{\label{Fig:jpgtttt} (a) Absorption intensity for different incident angles with TM-polarized wave. The electric field distributions with incident angle 20° at the two splitting modes (b) 1.59 THz and (c) 1.82 THz, where arrows show the directions of electric field distributions. (d) Schematic drawing of a photonic crystal slab with structural distortion into an ellipse. The length of the major and minor axes is denoted by $r_1$ and $r_2$, respectively. In this plot, the photonic crystal slab is simulated based on the optimized parameter of $\Lambda$ = 87 $\mu$m and thickness \emph{d} = 23 $\mu$m. (e) and (f) shows the transmission and the absorption intensity, respectively, as a function of frequency and $r_1$. Here $r_2$ is fixed as 30 $\mu$m. (g) to (i), same as (d) to (f) but with $r_1$ and $r_2$ interchanged.}
	\end{SCfigure*}

	The guided resonance in the photonic crystal slab exhibits signatures of Fano resonances \cite{fan2002analysis,zhou2019multiple,magnusson1995transmission}. 
	The transmission spectrum consists of asymmetric Fano line shape [Fig. 2(a)]. 
	In contrast, the absorption spectrum shows a \emph{symmetric} Lorentzian line shape, which is in accordance with the CMT description of the optical resonator [see Eqs. (A18)-(A26) in Appendix A]. 
	We also compare the FDTD simulation with the CMT calculation for the two resonances in Fig. 2(a). The FDTD simulated absorption spectrum exhibits good agreement with the CMT-based theoretical calculation with the fitting model parameters $\gamma_1=\delta_1 = 0.0634$ THz, $\gamma_2=\delta_2 = 0.0216$ THz, where $\gamma_i$ denotes the radiation rate and $\delta_i$ is the dissipative loss rate, and the subscript $i=1,2$ denotes the resonance peaks at 1.70 THz and 2.33 THz, respectively. 
	Notably, the fitted parameters of $\gamma=\delta$ for both absorption peaks confirm that the resonance effect originates from the critical coupling effect [see Appendix B for detailed discussion on the fitted $\gamma$ and $\delta$ from FDTD]. 
	%Furthermore, for multi-mode critical coupling, the radiation rates of different resonance modes are not necessarily to be the same. 
	The good agreement between FDTD and CMT obtained here suggests that CMT can be utilized as a simplified computational tool for the accelerated design of critical coupling effects in multi-mode resonance system. 
	
	The absorption peaks under incident lights of TE and TM polarizations at normal incidence is shown in Fig. 2(b). The optical response is almost identical for both the TE and TM polarizations, thus indicating the polarization-independent nature of the critical coupling effect. To understand the electric field distribution at the absorption peaks, 
	the electric field intensity distributions $|Ez|^2$ in the plane of the resonator (\emph{xy} cross-sectional plane through the center of the silicon slab) are presented in Figs. 2(c) and (d) for the absorption peaks at 1.7 THz and 2.33 THz, respectively. The electric field intensity for the 1.7 THz resonance peak concentrates at the edge of the air hole, with a substantial portion of electric fields at the high index position of the slab. In contrast, for the resonance peak at 2.33 THz, the electric field intensity distribution is strongly localized around edge of the hole.
	Despite the slightly different electric field distribution between the two resonance frequencies, Figs. 2(c) and (d) demonstrate that strongly localized electric field enhancement at the the edge of the hole in the photonic crystal slab is a common key feature in attaining strong optical absorption for the two resonance frequencies.

	\subsubsection{Oblique incidence and structural distortions}
	
	We now evaluate the robustness of the reconfigurability of the proposed device via two performance metrics, namely the optical response deviations at (i) oblique incidence; and with (ii) structural distortion of the photonic crystal.

	For (i), the device ideally should provide robust spectra intensity when in a wide range of incidence angle. To investigate this aspect, the absorption resonance peak located at 1.7 THz is simulated for a TM-polarized light with incident angle ranging from $0^\circ$ to $55^\circ$. Intriguingly, the resonance peak splits into two distinct modes at oblique angles as shown in Fig. 3(a). For the photonic crystal slab, the effective propagation constant of the guided mode resonance is 
	\begin{equation}
		k_{x,m} \cong \beta_{m}  = {k_0}\left( {{n_c}\sin {\varTheta_0} - m\frac{\lambda }{\Lambda }} \right), 
	\end{equation}
	where $m = \pm1, \pm2,\cdots$, $k_0= 2\pi/\lambda$ ($\lambda$ is incident wavelength) and $\varTheta_0$ is the incident angle. According to the guided mode resonance condition, the oblique incidence destroys the spatial symmetry of the wave, resulting in different magnetude of the propagation constant $|\beta_{m}|$ at oblique incidence and giving rise to two distinct splitting modes. 
	%This behavior can be readily explained by our simulations as the following. 
	For $\varTheta_0=5^\circ$, the magnitude of the absorption peak of the two splitting modes are about 0.98 and 0.89, respectively. When the angle is further increased, the absorption intensity gradually decreases. Nonetheless, the absorption remains at about 85\% for one of the splitting modes even for $\pm30^\circ$, thus suggesting the \emph{wide-angle} operation capability of the proposed device. 
	To further understand the two splitting resonances, we examine the distributions of the electric fields under an oblique incident angle of 20$^\circ$ [Figs. 3(b) and 3(c)]. The electric fields exhibit similar behaviors for both splitting modes and are anti-phased to each other. Such anti-phased electric field patterns confirm that the splitting modes originate from the spatial symmetry breaking of the wave order as mentioned above.

	\begin{figure*}[t]
		\includegraphics[width=1\linewidth]{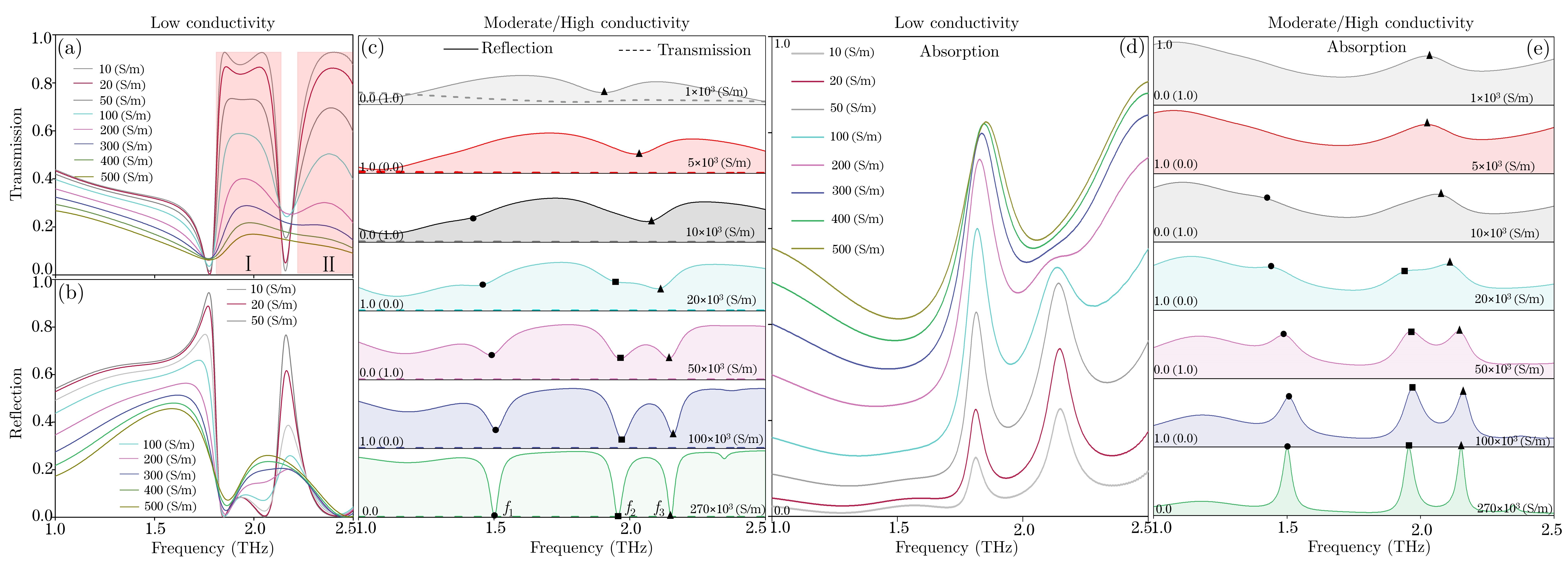}
		\label{Fig:jpg1}
		\caption{Simulated transmission (a) and reflection (b)-(c) spectra at a series of VO$_2$ conductivity ($\sigma$ in unit of S/m). In (c), vertical axis of each panel has a numerical range from 0 to 1. Simulated absorption spectra at  low conductivity state (d) and moderate to high conductivity state (e).}
	\end{figure*}
	
	For (ii), the effect of structural distortion on the absorption and the transmission spectra of the proposed device under normal incidence are investigated [see Figs. 3(d) to (i)]. 
	We consider the case where the circular holes in the photonic crystal slab are distorted into ellipses. Two configurations are investigated: the major axis of the ellipse is along the (A) y-axis [Fig. 3(d)]; and the (B) \emph{x}-axis [Fig. 3(g)] and the major axis radius is fixed at 30 $\mu$m. For configuration (A) in the insulating phase, high transmission intensity covers a wide frequency range as denoted by Regions \Rmnum{1} and \Rmnum{2} in Fig. 3(e). The high-transmission frequency bandwidth slightly decreases with an increasing $r_1$ in Region I. In contrast, Region II covers a broader range of frequency with an increasing $r_1$. However, for configuration (B), when $r_1$ is fixed, transmission intensity in both regions I and II show stable and high efficiency. Also, in this case, another high transmission region \Rmnum{3} exists when $r_2$ increases from 15 to 20 $\mu$m, and subsequently gradually diminishes [Fig. 3(h)]. These results demonstrated that the overall transmission peaks are not significantly affected by the structural perturbation.

	Comparing  the transmission spectra under fixed $r_1$ or $r_2$ condition, the absorption spectra display several important features. As shown in the absorption spectra for fixed $r_1$ and $r_2$ as a function of frequency [see Figs. 3(f) and 3(i),
	respectively], absorption peaks occur via three resonance modes, denoted as $f_1$, $f_2$ and $f_3$. These resonant frequencies exhibit a blue-shift with increasing $r_1$ or $r_2$. We attribute this blue-shift to the decrease of the effective refractive index of the photonic crystal slab arising from the perturbed structural parameters. 
	Figures 3(d)-3(i) thus suggest that the transmission and absorption peak responses of the proposed devices are robust against deviation from the perfect circular morphology of the photonic crystal. Importantly, the shifting of the resonance energies suggests that elliptical hole or other complex geometries could provide a tuning knob to engineer the transmission and absorption properties of the proposed device structures.

	\subsubsection{Transmission and absorption spectra evolution during VO$_2$ phase transition}

	To illustrate the intermediate state of the proposed PCM-based optical resonator and the evolution of the transmission, reflection and absorption spectra during VO$_2$ IMT, the optical response of the proposed device is simulated with VO$_2$ varying from 10 to 2.7 $\times$ 10$^5$ S/m. 
	Optimized device parameters of $\Lambda$ = 87 $\mu$m, thickness \emph{d} = 23 $\mu$m, and air hole radius \emph{r} = 30 $\mu$m are chosen, so to enable three resonance modes in the range of 1.0 to 2.5 THz.

	%Figures 4(a) and 4(b) illustrate the transmission and reflection spectra for low-conductivity regime of 10 S/m to 500 S/m, respectively, while Fig. 4(c) illustrate the reflection and transmission spectra at the moderate-to high-conductivity regime of $10^3$ to $10^5$ S/m. The absorption spectra of the low-conductivity and the moderate-to high-conductivity regime is shown in Figs. 4(c) and 4(d), respectively. 

	Figure 4 demonstrates the continuous tuning of the optical responses of the proposed devices as mediated by the IMT of the VO$_2$ slab. 
	When VO$_2$ is in the low conductivity state of 10 S/m, the VO$_2$ behaves as a dielectric substrate and the photonic crystal slab produces guided mode resonances with strong transmission peaks. A transmission peak as high 92\% can be achieved with the proposed structure [Fig. 4(a)]. Correspondingly, the reflection curve exhibits a minimum as denoted by the dark gray line in Fig. 4(b). Meanwhile, as the system is operating in the transmission mode, the lack of a metallic back mirror results in minimal absorption [see Fig. 4(c)]. As the conductivity of VO$_2$ is increased from 10 to 500 S/m, the transmission intensity in the shaded Regions I and II of Fig. 4(a) diminishes, while the reflection increases slightly [Fig. 4(b)]. On the other hand, absorption intensity gradually increases [see Fig. 4(d)], indicating a switching from the transmission to absorption. The transmission mode almost disappear when the conductivity reaches 5000 S/m [Fig. 4(c)]. 
	
	Notably, for VO$_2$ with moderate conductivity $10^3$ to $10^4$ S/m, the device exhibits broadband absorption well over 50\% over the entire 1 to 2.5 THz regime [Fig. 4(e)], which can be useful for broadband THz absorber application. 
	Further increasing the conductivity of VO$_2$ surpresses both transmission and reflection. With a VO$_2$ conductivity of 2.7 $\times$ $10^5$ S/m, both the transmission and reflection peaks are maximally suppressed [Fig. 4(c)], while the near perfect absorption peaks, as previously discussed Fig. 3, can be clearly observed [Fig. 4(e)].
	Experimentally, the phase transition of the VO$_2$ slab can be reversibly triggered by an electrical voltage \cite{shu2021electrically}. The proposed device is thus compatible with all-electric switching operation that can be directly integrated with digital electronic systems \cite{zhang2021electrically}. 
	
	Finally, we note that other than VO$_2$, other PCM (such as Ge$_2$Sb$_2$Te$_5$) has also been previously employed as a back mirror for reconfigurable metasurface applications \cite{carrillo2018reconfigurable,tittl2015switchable,carrillo2016design,cao2014broadband,galarreta2018nonvolatile}. However, it should be noted that in such devices, the PCM primarily serves as a back mirror or modulating substrate whose switchable optical properties transform the device between absorption and reflection modes under \emph{one-port} configuration throughout the device operation. In contrast, the \emph{port reconfigurable} optical resonator device proposed in this work dynamically switches between one-port absorption and two-port transmission modes when the the PCM substrate is switched between \emph{reflective back mirror} and \emph{transmissive back `glass'}, respectively. The proposed port reconfigurable optical resonator is thus radically different compared to the prior proposals of PCM-based reconfigurable device with \emph{fixed port configuration}.  
	
	\section{Conclusions}
	In conclusion, a design of phase change material-based port reconfigurable optical resonator is proposed. By incorporating VO$_2$ as a substrate under a photonic crystal slab, the dynamical switching of the port configuration from one-port to two-port configurations and the continuous evolution of the optical resonator between the transmission and the absorption modes is demonstrated using FDTD simulations combined with coupled mode theory analysis. Intriguingly, the port reconfigurable optical response of the proposed device remains robust against oblique incidence and structural perturbation, thus suggesting the robustness and practicality of the proposed device. 
	Our findings shall pave a radical strategy for designing highly efficient reconfigurable optical modulator based on the port reconfiguration mechanism uniquely enabled by phase change materials.

	\begin{acknowledgements}
		Y.S.A. is supported by the Singapore University of Technology and Design (SUTD) Start-up Research Grant (SRG SCI 2021 163). 
		%C.H.L. is supported by WBS:R-144-000-435-133. 
		H.Y.M. is supported by the China Scholarship Council, National Natural Science Foundation of China (61775055).
		L.W. acknowledges the support of National Natural Science Foundation of China (61775055).%XXXXXXXXXXXX and YYYYYYYYYYYY \textcolor{red}{[Haiyu, please include your funding, such as CSC and your Hunan university scholarship]}
		
	\end{acknowledgements}

	\appendix
	
	\section{Coupled mode theory}
	
	We develop the coupled model theory (CMT) model \cite{fan2003temporal,horng2020perfect,xiao2020tailoring,haus1984waves} in this Appendix, which is used to describe the input-output properties of a resonator. The generic geometry is illustrated in Fig. 5. The yellow area represents a single-mode optical resonator with amplitude \emph{a} coupled to $\mu$ ports, where power exchange takes place. The optical behavior for the resonance mode can be described by CMT equations. Detailed key results of the analysis are stated in follows.
	\begin{equation}
		\frac{da}{dt}=j\omega_0a-(\gamma+\delta)a+|\kappa\rangle| s_+\rangle
	\end{equation}
	\begin{equation}
		| s_-\rangle=C| s_+\rangle+a|\kappa\rangle^*
	\end{equation}
	In this coupling region, $\omega_0$ is the center frequency of resonance. The amplitude \emph{a} is normalized so that $|a|^2$ corresponds to the energy store inside the resonator. The decay rates $\gamma$ and $\delta$  represent two physically distinct mechanisms, namely the radiation rate and the dissipative loss rate of the model. The existence of $\gamma$ and $\delta$ indicates that the resonant process is not lossless and resonance decays into the ports with these two decay rates. From Eq. (A1), we can conclude that the resonator mode is described by three parameters: $\omega_0$, the two decay rates of amplitude caused by $\gamma$ and $\delta$. The amplitude \emph{a} is time dependence denoted with exp ($j\omega$\emph{t}), which is demonstrated by the term $\frac{da}{dt}=j\omega_0$\emph{a}. This resonance phenomenon represents a time dependence harmonic field. $|\kappa\rangle| s_+\rangle$ is introduced representing an electromagnetic wave excitation, which is an external signal. Except for this term, other terms represent the inherent properties of the resonator. $| s_+\rangle=[\emph{s}_{1+}, s_{2+}, \cdots, s_{\mu+}]^T, | s_-\rangle=[\emph{s}_{1-}, s_{2-}, \cdots, s_{\mu-}]^T, |\kappa\rangle=[\kappa_1, \kappa_2, \cdots, \kappa_\mu]^T, |\kappa\rangle^*$ represent the incoming and outgoing amplitude of plane waves, the coupling constants of resonant mode coupled with incoming and outgoing waves. 
	\begin{figure}
		\center{\includegraphics[width=8.8cm,height=3.6cm]{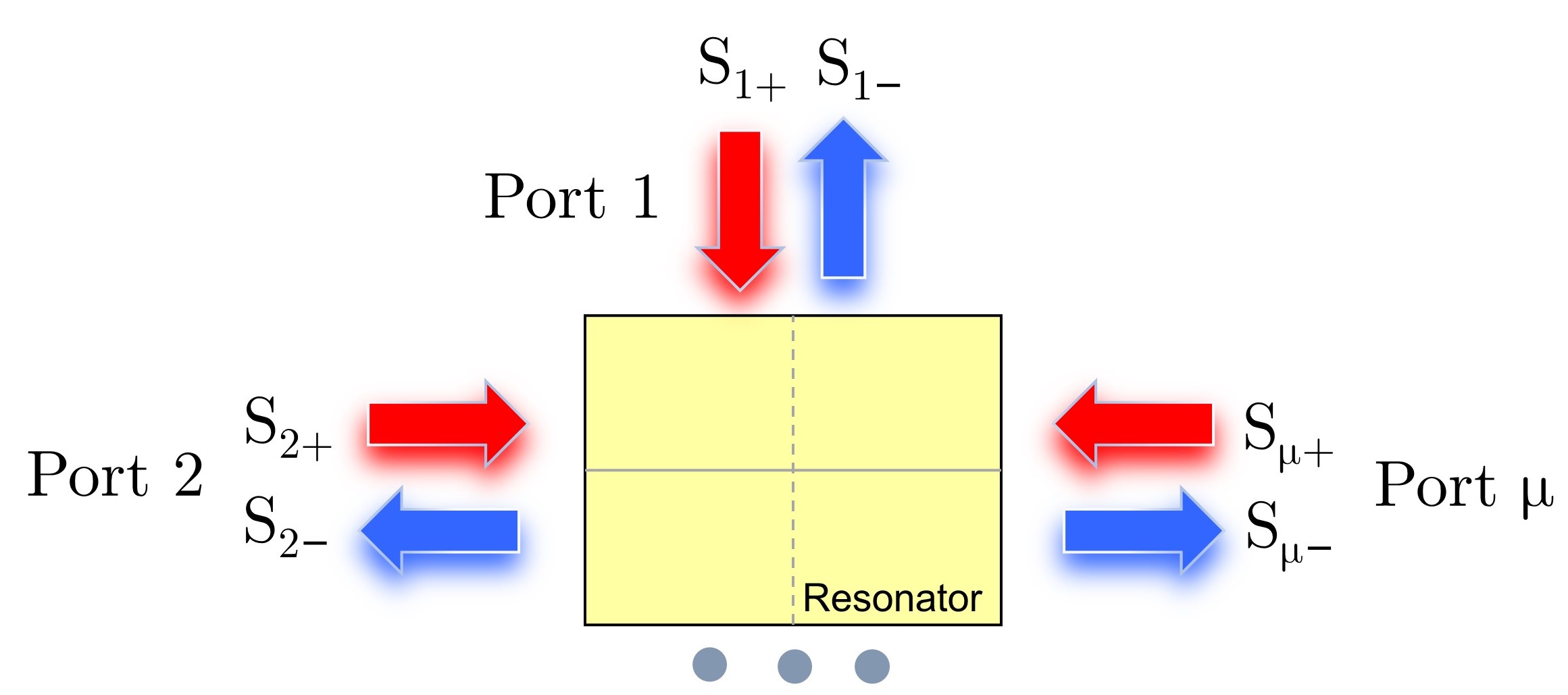}}
		\caption{\label{Fig:jpg} Schematic of coupling effect for an optical resonator system coupled to $\mu$ ports. The red and blue arrows indicate the incoming and outgoing waves, respectively. CMT can be used to describe the input-output properties of this resonator.}
	\end{figure}
	If the externally incident excitation  $| s_+\rangle$ is at frequency $\omega$ under the harmonic field $\frac{da}{dt}=j\omega a$, from Eq. (A1), the response can be written as
	\begin{equation}
		j\omega a=j\omega_0a-(\gamma+\delta)a+|\kappa\rangle| s_+\rangle  
	\end{equation}
	\begin{equation}
		j[(\omega-\omega_0)+(\gamma+\delta)]a=|\kappa\rangle| s_+\rangle
	\end{equation}
	\begin{equation}
		a = \frac{{\left| \kappa  \right\rangle \left| {{s_ + }} \right\rangle }}{{j(\omega  - {\omega _0}) + \gamma  + \delta }}
	\end{equation}
	The energy store $|a|^2$ inside the resonator can be written as
	\begin{equation}
		|a|^2=\frac{\left\langle\kappa|\kappa\right\rangle\left\langle s_+|s_+ \right\rangle}{(\omega-\omega_0)^2+(\gamma+\delta)^2} 
	\end{equation}
	From Eq. (A6), we note that the energy storage is independent of \emph{C} which gives the direct transmission and reflection coefficients. The coefficient  $|\kappa\rangle$ is related to energy conservation and time-reversal symmetry constraints. First, we apply time-reversal symmetry to the case with $\delta=0$  (with no internal loss) and consider the situation in which there is no external excitation $| s_+\rangle=0$ . Then, according to Eq. (A1), we can obtain 
	\begin{equation}
		\frac{da}{dt}=j\omega_0a-\gamma a
	\end{equation}
	Second, according to energy conservation and Eq. (A7), in the scenario where the external excitation is absence, the energy stored inside the resonator is taken away by the outgoing waves. Thus, the time rate of change of the energy can be written as 
	\begin{equation}
		\frac{d|a|^2}{dt}=a^*\frac{da}{dt}+a\frac{da^*}{dt}=-2\gamma |a|^2=-\langle s_-|s_-\rangle
	\end{equation}
	Eq. (A8) indicates that $|a|^2$  decays as exp ($-2\gamma t$).

	\begin{figure}
		\center{\includegraphics[width=8.8cm,height=8.5cm]{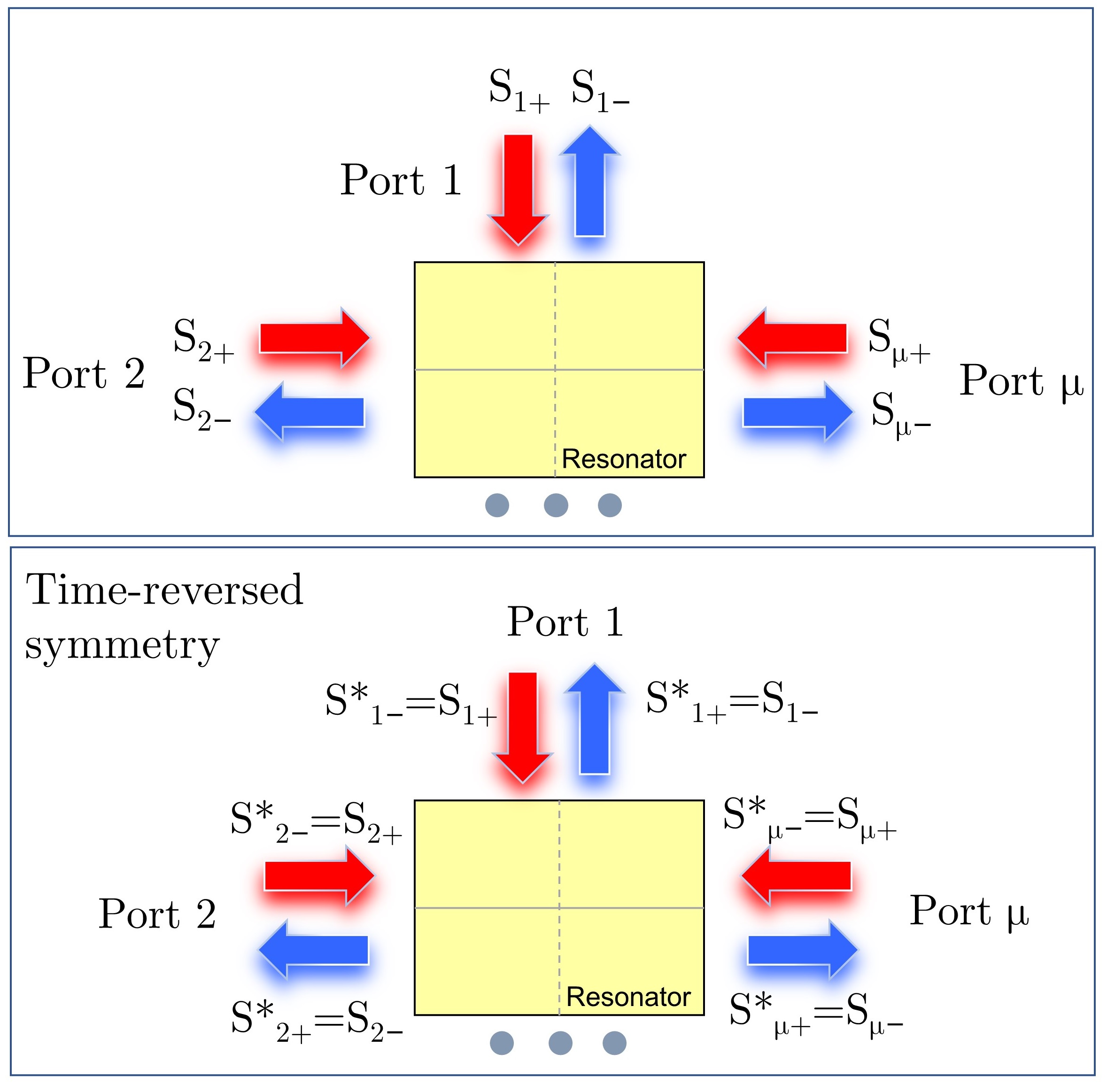}}
		\caption{\label{Fig:jpg} Schematic of coupling effect for an optical resonator system coupled to $\mu$ ports under time reversed condition.}
	\end{figure}
	
	Through the time-reversed solution, the process of energy exchange in the resonator describes that the wave energy in the resonator buildups, rather than decays, with the time dependence exp ($+2\gamma t$)  rather than ($-2\gamma t$). Meanwhile, the wave is outgoing rather than incoming. Here, we interpret $|\ s_-\rangle^*$ as the time-reversed amplitudes of the input wave as shown in Fig. 6. Combing the above terms, we obtain that $|\ s_+\rangle \rightarrow|\ s_-\rangle^*$, $|\ s_-\rangle \rightarrow|\ s_+\rangle^*$, $a_+(t)$ dependence exp ($j\omega t$) $\rightarrow$ $a_-(t)$ dependence exp ($-j\omega t$). The time-reversed case is represented by the incident wave $|\ s_-\rangle^*$ at frequency $\omega_0$, and grows at the rate and a complex frequency  $\omega=\omega_0-\gamma j$ \cite{fan2003temporal,haus1984waves}. Introducing this frequency into Eq. (A5) with $\delta=0$, we obtain that 
	$$a = \frac{{\left| \kappa  \right\rangle \left| {{s_ + }} \right\rangle }}{{2\gamma }}$$
	\begin{equation}
		{a^*} = \frac{{{{\left| \kappa  \right\rangle }^*}{{\left| {{s_ + }} \right\rangle }^*}}}{{2\gamma }} = \frac{{{{\left| \kappa  \right\rangle }^*}\left| {{s_ - }} \right\rangle }}{{2\gamma }}
	\end{equation}
	According to Eq.(A8) and (A9),
	\begin{equation}
		\left\langle\ s_-|\ s_-\right\rangle=2\gamma|a|^2
	\end{equation}
	By applying the time-reversed and combing Eqs. (A8)-(A10), it allows us to calculate the coupling coefficient as
	\begin{equation}
		\langle\kappa|\kappa\rangle=2\gamma, |\kappa|=\sqrt{2\gamma}
	\end{equation}

	\begin{figure*}[t]
		\center{\includegraphics[width=1\linewidth]{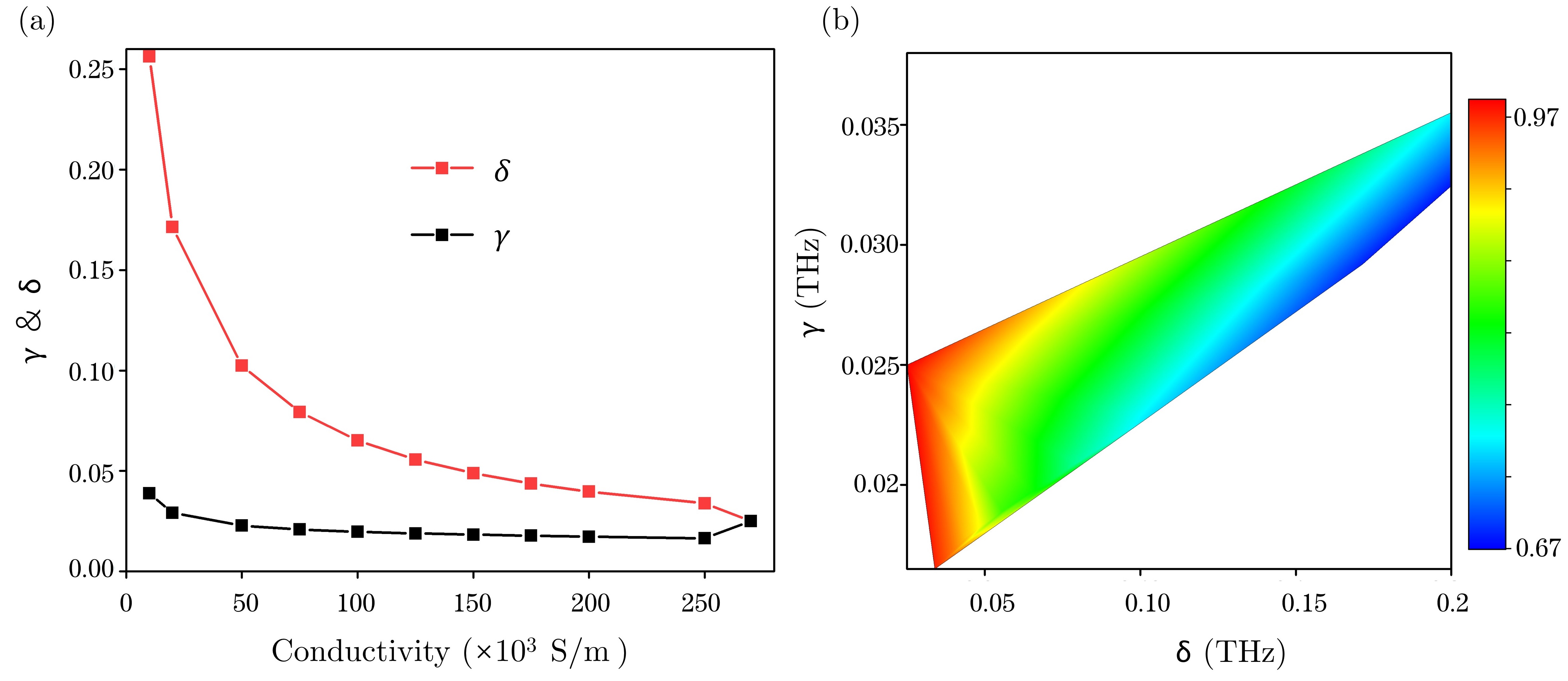}}
		\caption{\label{Fig:jpg} Calculated coupling coefficient $\gamma$ and $\delta$ using CMT as a function of the conductivity of VO$_2$. In this plot, we chose the second resonant mode of Fig. 2(a) to illustrate this relationship. (b) Absorptance diagrams for the chosen mode varies against $\gamma$ and $\delta$.}
	\end{figure*}
	
	After the coupling constant is solved, the coefficient \emph{C} can be calculated by energy conservation which requires that the net power flowing into the resonator is equal to the rate of buildup of energy within the resonator plus the rate of energy dissipation:
	\begin{equation}
		\langle s_+|s_+\rangle-\langle s_-|s_-\rangle=\frac{d|a|^2}{dt}+2\delta|a|^2
	\end{equation}
	On the other hand, from Eq. (A1), we find 
	\begin{equation}
		\begin{aligned}
			&& \frac{d|a|^2}{dt} &=a^*\frac{da}{dt}+a\frac{da^*}{dt}=a^*[j\omega_0a-(\gamma+\delta)a+|\kappa\rangle| s_+\rangle]\\
			&& \ & +a[-j\omega_0a^*-(\gamma+\delta)a^*+|\kappa\rangle^*| s_+\rangle^*]\\
			&& \ & =-2(\gamma+\delta)|a|^2+(a^*|\kappa\rangle| s_+\rangle+a|\kappa\rangle^*| s_+\rangle^*)
		\end{aligned}
	\end{equation}
	\begin{equation}
		\langle s_+|s_+\rangle-\langle s_-|s_-\rangle=-2\gamma|a|^2+(a^*|\kappa\rangle| s_+\rangle+a|\kappa\rangle^*| s_+\rangle^*)
	\end{equation}
	\begin{equation*}
		(1-|C|^2)\langle s_+|s_+\rangle+(-|C|^*-1)| s_+\rangle^*|\kappa\rangle^*a+
	\end{equation*}
	\begin{equation}
		(-C-1)a^*|\kappa\rangle| s_+\rangle=0
	\end{equation}
	Combining Eqs.(A2),(A12)-(A15), the energy conservation allows us calculate the coefficient \emph{C} as
	\begin{equation}
		\begin{cases}
			(1-|C|^2)=0, & \ C^*C=1\\
			(-C^*-1)=0, & \ C^*=-1\\
			(-C-1)=0, & \ C=-1
		\end{cases}
	\end{equation}
	Thus, Eq. (A2) can be written as $| s_-\rangle=-| s_+\rangle+a|\kappa\rangle^*$.
	The output $| s_-\rangle$ can be described directly using the scattering matrix \emph{S} of the system for externally incident $| s_+\rangle$, $| s_-\rangle=\emph{S}| s_+\rangle$ \cite{xiao2020tailoring}. Assuming $\delta=0$, 
	\begin{equation}
		\begin{aligned}
			S &=exp(i\phi)\begin{bmatrix}
				r_d & \ & jt_d \\
				jt_d & \ &  r_d
			\end{bmatrix}\\
			& +\frac{\gamma}{j(\omega-\omega_0)+\gamma}
			\begin{bmatrix}
				-(r_d\pm jt_d)     & \  & \mp (r_d\pm jt_d) \\
				\mp (r_d\pm jt_d)  & \  &  -(r_d\pm jt_d)
			\end{bmatrix}
		\end{aligned}
	\end{equation}
	where $r_d$  and $t_d$  are the direct reflection and transmission coefficients with $t_d=\sqrt{1-r_d^2}$. $r_d$, $t_d$ and  $\phi$ are all real constant. The plus and minus sign in Eq. (A17) represent the even and odd resonance modes, respectively. The decaying amplitudes to the side of the slab are in phase and 180° out of phase for even and odd modes, respectively. The intensity reflection coefficient $R=|S_{11}|^2$ and transmission coefficient $T=|S_{21}|^2$ are derived as
	\begin{equation}
		R=\frac{r_d^2(\omega-\omega_0)^2+t_d^2\gamma^2\mp2r_dt_d(\omega-\omega_0)\gamma}{(\omega-\omega_0)^2+\gamma^2}
	\end{equation}
	\begin{equation}
		T=\frac{t_d^2(\omega-\omega_0)^2+r_d^2\gamma^2\pm2r_dt_d(\omega-\omega_0)\gamma}{(\omega-\omega_0)^2+\gamma^2}
	\end{equation}
	From Eqs. (A18) and (A19), we note that, when $t_d=1$, $r_d=0$ or $t_d=0$, $r_d=1$ the intensity reflection and transmission coefficient become 
	\begin{equation}
		R=\frac{\gamma^2}{(\omega-\omega_0)^2+\gamma^2}
	\end{equation}
	\begin{equation}
		T=\frac{(\omega-\omega_0)^2}{(\omega-\omega_0)^2+\gamma^2}
	\end{equation}
	or
	\begin{equation}
		R=\frac{(\omega-\omega_0)^2}{(\omega-\omega_0)^2+\gamma^2}
	\end{equation}
	\begin{equation}
		T=\frac{\gamma^2}{(\omega-\omega_0)^2+\gamma^2}
	\end{equation}
	Thus, a symmetric Lorentzian line shape is produced when either  $r_d$ or $t_d$ is zero. In the general case  $ r_d\ne 0$ or $t_d\ne 0$, the transmission and reflection spectra exhibit a Fano asymmetric line shape, while the absorption spectra will be a sum of symmetric Lorentzian line shape.
	
	For a one-port system, the back reflector prohibits any transmission through the system so that only the reflection needs to be considered. According to Eq. (A2) and (A4), the reflection coefficient  can be obtained.
	\begin{equation}
		\left| {{s_-}}\right\rangle = -\left| {{s_ + }} \right\rangle  + \frac{{\left\langle {\kappa }\mathrel{\left|{\vphantom{\kappa\kappa}}\right.\kern-\nulldelimiterspace}{\kappa}\right\rangle\left|{{s_ + }} \right\rangle }}{{j\left( {\omega  - {\omega _0}} \right) + \gamma  + \delta }}
	\end{equation}
	%\begin{equation}\label{T}
	%\Gamma=\frac{|\right S_-\rangle}{|\right S_+\rangle}
	%=\frac{2\gamma}{j(\omega-\omega_0)+\gamma+\delta}-1=\frac{-j(\omega-\omega_0)+\gamma-\delta}{j(\omega-\omega_0)+\gamma+\delta}
	%\begin{array}{l}
	%\begin{aligned}
	%&& \Gamma  &= \frac{{\left| {{s_ - }} \right\rangle }}{{\left| {{s_ + }} \right\rangle }} = \frac{{2\gamma }}{{j\left( {\omega  - {\omega _0}} \right) + \gamma  + \delta }} - 1\\
	%&& {\rm{  }} &= \frac{{ - j\left( {\omega  - {\omega _0}} \right) + \gamma  - \delta }}{{j\left( {\omega  - {\omega _0}} \right) + \gamma  + \delta }}
	%\end{aligned}
	%\end{array}
	%\end{equation}
	
	\begin{equation}
		\Gamma=\frac{\langle s_-|s_-\rangle}{\langle s_+|s_+\rangle}
	\end{equation}
	
	Eqs. (A24) and (A25) show that the physical property of this model is determined by $\gamma$ and $\delta$. For a one-port structure when the system is driven on resonance ($\omega=\omega_0$) and the real and imaginary of the numerator equal to zero, $\Gamma$ vanishes. All incident power is absorbed in the system, meaning that critical coupling is achieved. As a result, the absorption at any frequency point of the spectral line for the system finally become 
	\begin{equation}
		\begin{array}{l}
			\begin{aligned}
				&& A &=1-\Gamma^2=\frac{\left\langle s_+|s_+\right\rangle-\left\langle s_-|s_-\right\rangle}{\left\langle s_+|s_+\right\rangle}\\
				&& &=\frac{4\gamma\delta}{(\omega-\omega_0)^2+(\gamma+\delta)^2}
			\end{aligned}
		\end{array}
	\end{equation}
	From the theoretical expressions above, the light absorption performance of the system is determined by the radiation loss and the dissipative loss. It is worth noting that the maximum absorption efficiency of a single optical resonator is highly related to its port numbers, as $A_{max}=\frac{1}{\mu}$. Therefore, for a given resonator, the port number is known, and then, $A_{max}$ can be determined. Generally, this theory developed only works for system with single resonant mode. For a system with exceed two resonances modes, this single mode CMT can be extended to multi-mode model and the reflection coefficient is written as \cite{qu2015tailor}
	\begin{equation}
		\Gamma  =  - 1 + \sum\nolimits_m {\frac{{2{\gamma _m}}}{{j\left( {\omega  - {\omega _m}} \right) + {\gamma _m} + {\delta _m}}}}
	\end{equation}
	where $\omega_m$, $\gamma_m$ and $\delta_m$ denote the \emph{m}-th resonance frequency, radiation rate and the dissipative loss rate of the \emph{m}-th mode, respectively. This multi-mode critical coupling theory allows us to calculate the reflection properties of a one-port resonator.

	\section{Comparing the CMT model and the FDTD simulations}
	
	In this Appendix, we demonstrate the relation between the critical coupling effect and the conductivity of VO$_2$ by examining the CMT coupling coefficients extracted from the FDTD simulation results. We plot the coupling coefficients $\gamma$ and $\delta$ as a function of the conductivity of VO$_2$ as shown in Fig. 7(a). With an increasing conductivity from 10 $\times$ $10^3$ to 2.7 $\times$ $10^5$ S/m, the coupling coefficient $\gamma$ and $\delta$ obtained from the analytical fitting successively decrease and converge to each others, indicating that by adjusting the conductivity of VO$_2$ the critical coupling condition of $\gamma = \delta$ is achieved, in agreement with the CMT model discussion presented in Appendix A [i.e. see Eqs. (A26) and (A27)]. 
	
	It can also be seen from Fig. 7(a) that the effect of VO$_2$ conductivity on $\gamma$ is rather weak, while it has a much stronger effect on $\delta$. 
	%In the scenario of $\gamma \neq \delta$ $\Gamma$ is not the minimum value, which set does not satisfy the critical coupling condition. 
	%From this calculation, it is easily concluded that the conductivity of VO$_2$ can significantly affect $\delta$, but it has relatively minor effect on $\gamma$. 
	From the CMT perspective, the conductivity of VO$_2$ thus play the role of alternating the $\gamma$ and $\delta$ coupling coefficient strengths in achieving the efficient modulation of the device transmission and absorption properties, as demonstrated in Fig. 4. 
	%Such relationships can help understand the different design of reconfiguration resonator assisted with the PCM. 

	In Fig. 7(b), the calculated $\gamma$ and $\delta$ in Fig. 7(a) are used to plot the map of absorption intensity versus $\gamma$ and $\delta$. Near-perfect absorption occur for the case $\gamma$ = $\delta$ = 0.0216, which is in consistency with the results in Fig. 2(a). The absorption intensity gradually decreases when the two parameters deviate apart from each other. Figure 7 thus suggests that the CMT model can be employed to adequately describe the behavior of the FDTD simulated device optical response through the coupling coefficients $\gamma$ and $\delta$ developed and discussed in Appendix A.

	\bibliographystyle{apsrev4-2}
	%\bibliography{Reference}
	

\end{document}